\documentclass[twocolumn,english,superscriptaddress,notitlepage]{revtex4-1}
\usepackage[T1]{fontenc}
\usepackage{graphicx}

\makeatletter
\usepackage{babel}

\makeatother

\begin{document}

\title{Reentrant Phase Diagram of $\rm{Yb_2Ti_2O_7}$ in $\langle 111 \rangle$ Magnetic Field}

\author{A. Scheie}
\address{Department of Physics and Astronomy, Johns Hopkins University, Baltimore, MD 21218}
\address{Institute for Quantum Matter, Johns Hopkins University, Baltimore, MD 21218}

\author{J. Kindervater}
\address{Department of Physics and Astronomy, Johns Hopkins University, Baltimore, MD 21218}
\address{Institute for Quantum Matter, Johns Hopkins University, Baltimore, MD 21218}

\author{S. S\"aubert}
\affiliation{Physik-Department, Technische Universit{\"a}t M{\"u}nchen, D-85748 Garching, Germany}
\affiliation{Heinz Maier-Leibnitz Zentrum, Technische Universit{\"a}t M{\"u}nchen, D-85748 Garching, Germany}

\author{C. Duvinage}
\affiliation{Physik-Department, Technische Universit{\"a}t M{\"u}nchen, D-85748 Garching, Germany}

\author{C. Pfleiderer}
\affiliation{Physik-Department, Technische Universit{\"a}t M{\"u}nchen, D-85748 Garching, Germany}

\author{H. J. Changlani}
\address{Department of Physics and Astronomy, Johns Hopkins University, Baltimore, MD 21218}
\address{Institute for Quantum Matter, Johns Hopkins University, Baltimore, MD 21218}

\author{S. Zhang}
\address{Department of Physics and Astronomy, Johns Hopkins University, Baltimore, MD 21218}
\address{Institute for Quantum Matter, Johns Hopkins University, Baltimore, MD 21218}

\author{L. Harriger}
\address{NIST Center for Neutron Research, National Institute of Standards and Technology, Gaithersburg, MD 20899}

\author{K. Arpino}
\address{Department of Chemistry, Johns Hopkins University, Baltimore, MD 21218}
\address{Institute for Quantum Matter, Johns Hopkins University, Baltimore, MD 21218}

\author{S.M. Koohpayeh}
\address{Department of Physics and Astronomy, Johns Hopkins University, Baltimore, MD 21218}
\address{Institute for Quantum Matter, Johns Hopkins University, Baltimore, MD 21218}

\author{O. Tchernyshyov}
\address{Department of Physics and Astronomy, Johns Hopkins University, Baltimore, MD 21218}
\address{Institute for Quantum Matter, Johns Hopkins University, Baltimore, MD 21218}

\author{C. Broholm}
\address{Department of Physics and Astronomy, Johns Hopkins University, Baltimore, MD 21218}
\address{Institute for Quantum Matter, Johns Hopkins University, Baltimore, MD 21218}
\address{NIST Center for Neutron Research, National Institute of Standards and Technology, Gaithersburg, MD 20899}
\address{Department of Materials Science and Engineering, Johns Hopkins University, Baltimore, MD 21218}

\date{\today}

\begin{abstract}
We present a magnetic phase diagram of rare-earth pyrochlore $\rm{Yb_2Ti_2O_7}$ in a $\langle 111 \rangle$ magnetic field. Using heat capacity, magnetization, and neutron scattering data, we show an unusual field-dependence of a first-order phase boundary, wherein a small applied field increases the ordering temperature. The zero-field ground state has ferromagnetic domains, while the spins polarize along $\langle 111 \rangle$ above 0.65T. A classical Monte Carlo analysis of published Hamiltonians does account for the critical field in the low T limit. However, this analysis fails to account for the large bulge in the reentrant phase diagram, suggesting that either long-range interactions or quantum fluctuations govern low field properties.


\end{abstract}

\maketitle

$\rm{Yb_2Ti_2O_7}$ may be one of the most famous materials in frustrated magnetism, and yet its ground state has not been fully established. $\rm{Yb}^{3+}$ ions, each forming a Kramers doublet, occupy the vertices of a (pyrochlore) lattice of corner-sharing tetrahedra which frustrates the development of conventional long range order \cite{Hodges2002, HodgesCEF, GaudetRoss_CEF}. 
Much of the recent attention to $\rm{Yb_2Ti_2O_7}$ has been driven by the suggestion that this material forms a quantum spin-ice at low temperatures \cite{Ross_Hamiltonian, Hayre2013, ApplegateSpinIce, Armitage_monopoles_2016, MonopolesConductivity}, wherein the spins are constrained to point into or out of tetrahedra with a two-in-two-out "ice rule". This exotic state of matter is predicted to have a spin-liquid ground state with its own effective field theory \cite{SpinIce_review, Balents2010review}. The quantum spin ice (QSI) hypothesis is supported by evidence of monopoles in the paramagnetic phase \cite{ApplegateSpinIce, Armitage_monopoles_2016, MonopolesConductivity}, and diffuse zero-field inelastic neutron scattering \cite{GaudetRoss_order, Coldea2017}. Challenging the QSI hypothesis, however, is evidence that stoichiometric $\rm{Yb_2Ti_2O_7}$ ferromagnetically orders around 270mK (though the specific ordered structure is contested) \cite{FM_order2003, GaudetRoss_order, Yaouanc_order} with magnetic order enhanced under pressure \cite{Kermarrec2017}. It is unclear how to reconcile the ground state order of $\rm{Yb_2Ti_2O_7}$ with its more unusual behavior, especially since the ground state is not fully understood. What is more, there is limited experimental information about collective properties of $\rm{Yb_2Ti_2O_7}$ due to the lack of stoichiometrically pure crystals.



Here we report the phase diagram of stoichiometric $\rm{Yb_2Ti_2O_7}$ in a $\langle 111 \rangle$ magnetic field. The $\langle 111 \rangle$ field 
in pyrochlore compounds like $\rm{Yb_2Ti_2O_7}$ harbors the possibility of a quantum kagome ice phase \cite{KagIce_theory}; but our data does not reveal such a phase in $\rm{Yb_2Ti_2O_7}$. Instead, we find a reentrant phase diagram where magnetic order is enhanced under small magnetic fields--a behavior that extant models of $\rm{Yb_2Ti_2O_7}$ fail to explain when quantum fluctuations are neglected.

An unfortunate obstacle to studying $\rm{Yb_2Ti_2O_7}$ is that most single crystals are plagued by site disordered "stuffing", which causes large variations in the critical temperature \cite{SampleDependence_HC, SampleDependence_Ross, Ross_stuffing,DOrtenzio_noGsOrder}. This extreme sensitivity to disorder makes it difficult to compare experimental results to each other or to theory. 
Recently, however, high-quality stoichiometric single crystals were successfully grown with the traveling solvent floating zone method \cite{SeyedPaper}. We report the first field-dependent measurements on stoichiometric single crystals of $\rm{Yb_2Ti_2O_7}$, and we use them to build a phase diagram of $\rm{Yb_2Ti_2O_7}$ in a $\langle 111 \rangle$ magnetic field. In our analysis, we used three experimental methods: heat capacity, magnetization, and neutron scattering.

\begin{figure}
\centering\includegraphics[scale=0.55]{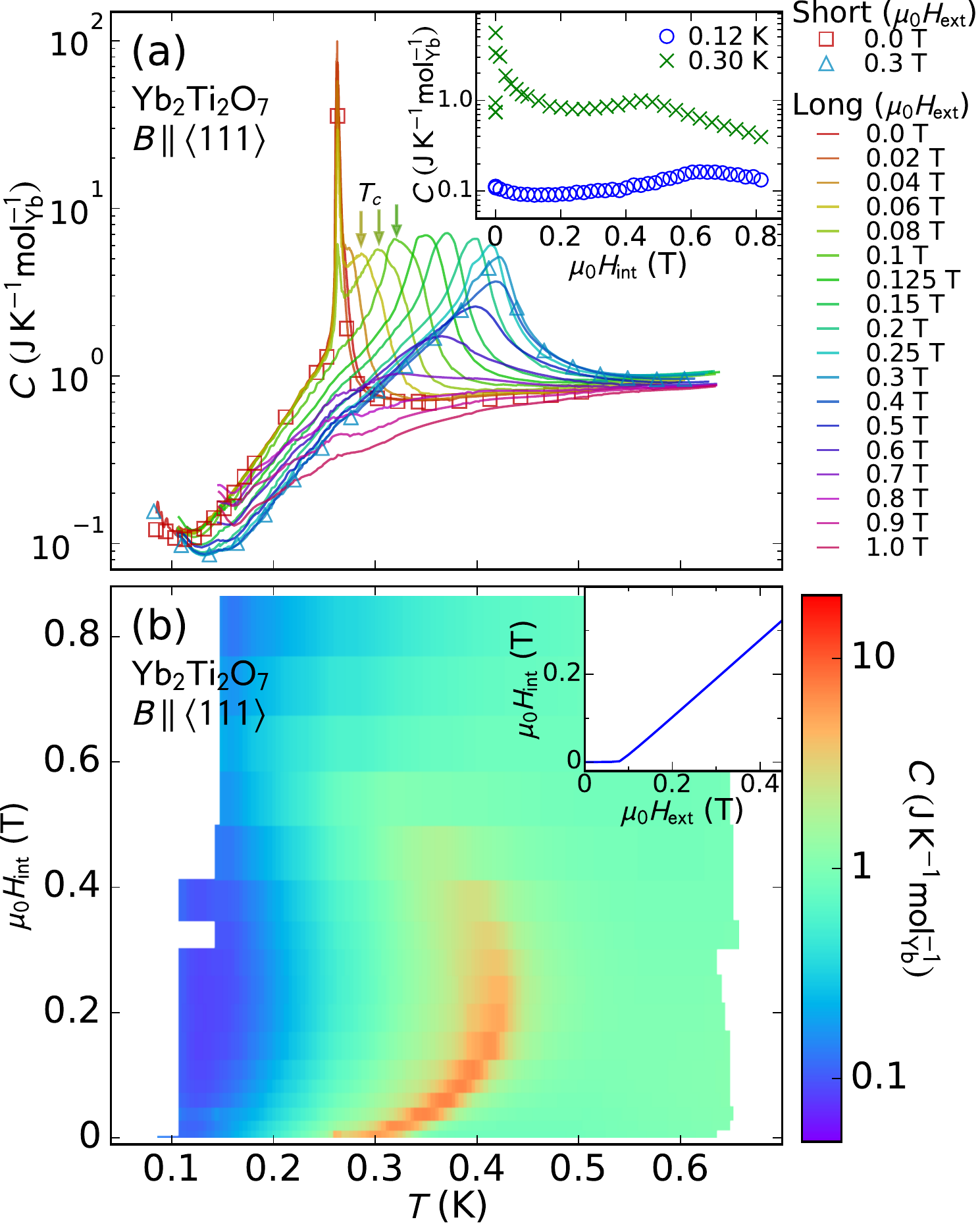}

\caption{Heat capacity data for $\mathrm{Yb_2Ti_2O_7}$ with magnetic field along $\langle 111 \rangle$. (a) $C$ vs. $T$ at various fields. Solid traces indicate long-pulse data, while discrete symbols indicate adiabatic short-pulse data. The fields in the legend are external fields. (a, inset) Isothermal field scans of heat capacity using the short-pulse method. (b) Color map of long-pulse heat capacity data vs. temperature and internal magnetic field. (b, inset) Relationship between $\mu_0H_{\rm int}$ and $\mu_0H_{\rm ext}$.}

\label{flo:HeatCapacity}
\end{figure}

The heat capacity of $\rm{Yb_2Ti_2O_7}$ at various magnetic fields is shown in Fig.~\ref{flo:HeatCapacity}.
We collected heat capacity data on a 1.04 mg sample of $\rm{Yb_2Ti_2O_7}$ in a $\langle 111 \rangle$ oriented magnetic field using a dilution unit insert of a Quantum Design PPMS \cite{NIST_disclaimer}. 
The heat capacity data were collected mostly with a long-pulse method, in which we applied a long heat pulse, tracked sample temperature as the sample cooled, and computed heat capacity from the time derivative of sample temperature (see ref. \cite{MethodsPaper} and supplemental materials for more details). 
The advantage of the long-pulse method is sensitivity to first order transitions, which  $\mathrm{Yb_2Ti_2O_7}$ is reported to have \cite{FirstOrderTransition,FirstOrder_magnetization,Chang}. Some adiabatic short-pulse data were taken as well, and  Fig.~\ref{flo:HeatCapacity} shows the overall agreement between these two methods.

The magnetic fields in Fig.~\ref{flo:HeatCapacity}(b) have been corrected for the internal demagnetizing field. The  demagnetization correction (see inset) is ${\bf H}_{int}={\bf H}_{ext}-D{\bf M}({\bf H}_{int})$, where $D$ is the demagnetization factor (determined by sample geometry), and ${\bf M}({\bf H}_{int})$ is magnetization (measured separately--see below). This correction enables quantitative comparison between measurements on differently shaped samples.


\begin{figure*}
\includegraphics[scale=1.0]{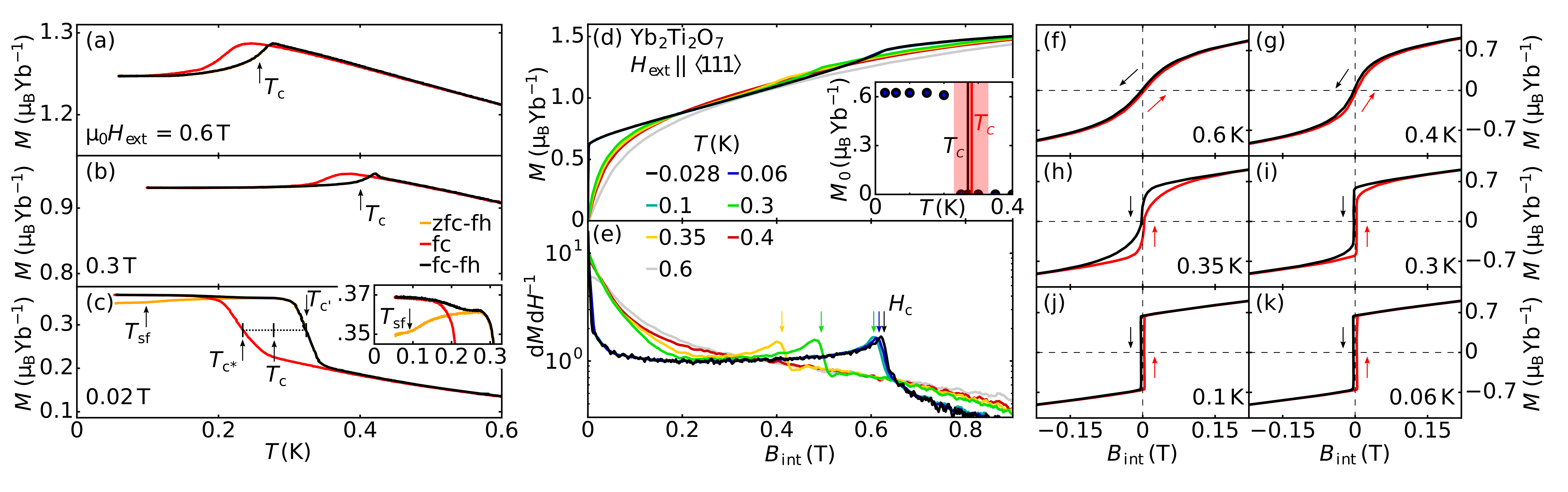}

\caption{Magnetization of  $\rm{Yb_2Ti_2O_7}$ in applied magnetic fields along $\langle 111 \rangle$. (a-c) Temperature dependence of the magnetization where we distinguish between data recorded according to procedure (i) zero-field-cooled / field-heated (zfc-fh), (ii) field-cooled (fc) and (iii) field-cooled / field-heated (fc-fh). (d) Magnetization and (e) numerical derivative of the experimental data of  $\rm{Yb_2Ti_2O_7}$ as function of internal magnetic field after correction for demagnetization fields. The inset in (d) shows the spontaneous magnetization as a function of temperature obtained from magnetization field sweeps (protocol (A1); see supplemental material). (f-k) Magnetization in field cycles of sweep types protocol (A2) and (A3) (see supplemental material).}

\label{flo:Magnetization}
\end{figure*}

The magnetization of  $\rm{Yb_2Ti_2O_7}$ (Fig. \ref{flo:Magnetization}) was measured by means of a bespoke vibrating coil magnetometer (VCM) as combined with a TL400 Oxford Instruments top-loading dilution refrigerator \cite{Legl2010a,Krey2012,Legl2012, NIST_disclaimer}. We measured the temperature dependence of the magnetization while cooling and while heating, with field-heating measurements performed on both a zero-field-cooled and a field-cooled state. Similarly, we measured field dependence magnetization with field sweeps from $0\rightarrow\,$1T performed on a zero-field-cooled sample, followed by field sweeps from +1T$\,\rightarrow\,$-1T and -1T$\,\rightarrow\,$+1T. Further details are provided in the supplementary material.
All magnetization measurements were carried out on a 4.7 mm diameter, 0.40 g sphere of $\rm{Yb_2Ti_2O_7}$, which was ground from a larger stoichiometric single crystal and polished into a spherical shape. The spherical geometry ensures a uniform demagnetization factor of $D=1/3$. 

\begin{figure}
\centering\includegraphics[scale=0.58]{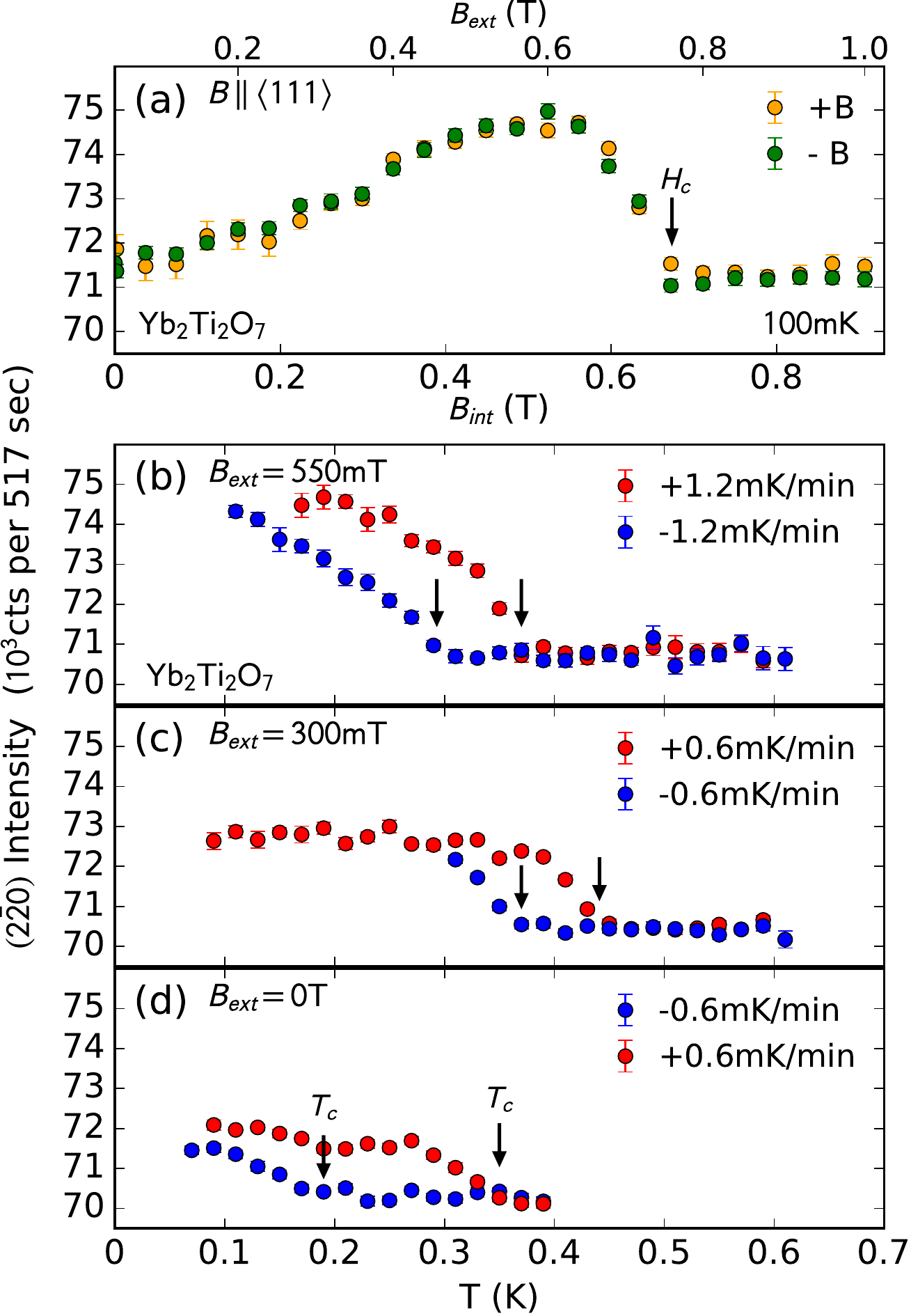}

\caption{Field and temperature dependence of the $(2\bar{2}0)$ Bragg peak intensity. (a) Magnetic field scan going up (+B) and down (-B) in field. Note the lack of hysteresis. (b-d) Temperature scans at external magnetic fields of 550mT, 300mT, and 0mT (internal fields of 473mT, 239mT, and 0mT). Red indicates increasing temperature, blue indicates decreasing temperature. Error bars indicate one standard deviation above and one standard deviation below the measured value.}

\label{flo:NeutronData}
\end{figure}

Finally, we collected neutron diffraction data at the SPINS cold neutron triple axis spectrometer at the NCNR. Our sample for these experiments was a 4.7mm $\rm{Yb_2Ti_2O_7}$ sphere (ground from the same crystal as the magnetization sample) in a dilution refrigerator with the $\langle 111 \rangle$ direction perpendicular to the scattering plane and along a vertical magnetic field, with $E_i=E_f=5 \> {\rm meV}$ neutrons and a full width at half maximum incoherent elastic energy resolution of $0.23 \> {\rm meV}$. To explore the phase boundaries seen in the heat capacity and magnetization measurements, we focused our attention on the $(2\bar20)$ peak, which was reported to be magnetic \cite{GaudetRoss_order, Yaouanc_order}. We first allowed the sample to settle into the ground state at zero field by cooling from 300 K over 17 hours and allowing the sample to sit for an additional seven hours at 65 mK. Following this, we scanned the applied magnetic field at 100 mK from 0 T to 1 T, and then performed slow temperature scans at various fields.  The results are shown in Fig. \ref{flo:NeutronData}. The neutron scattering measurements were taken with the detector sitting at the $(2\bar20)$ Bragg peak's maximum intensity, with periodic rocking scans to ensure alignment after cryogenic operations.

All three methods--heat capacity, magnetization, and neutron diffraction--point to a reentrant phase diagram of $\rm{Yb_2Ti_2O_7}$ in a $\langle 111 \rangle$ magnetic field.
The heat capacity plot in Fig.~\ref{flo:HeatCapacity} presents the clearest manifestation. At zero field, $C(T)$ has a sharp peak at 270 mK, as reported for stoichiometric powders  \cite{SampleDependence_Ross, SeyedPaper}, indicating a phase transition. A magnetic field initially causes this phase boundary to shift up in temperature, reaching a maximum temperature of 0.42K at an internal field of 0.24~T. At higher fields, the phase boundary sweeps back to 0K at 0.65~T. 
The data in Fig.~\ref{flo:HeatCapacity}(a) shows two heat capacity peaks at fields between 0.02 and 0.1 T. This is consistent with the result of field inhomogeneity from nonuniform demagnetizing fields in the plate-like specific heat sample. In a weak external field (below 0.1 T), the center of the sample still has no net internal field, giving rise to a residual sharp peak with the same $T_c$ as in zero external field. The residual peak disappears as soon as the entire sample has a non-zero net field. This field inhomogeneity would also broaden the peak in finite fields (see supplemental materials for more details).

The magnetization data in Fig. \ref{flo:Magnetization} contain several important features. First, Fig. \ref{flo:Magnetization}(a-e) confirm the reentrant phase diagram: the kinks and changes of slope in magnetization follow the same curved shape as the anomalies in heat capacity. The derivative $\mathrm{d}M/\mathrm{d}B$ shown in Fig.~\ref{flo:Magnetization}(e) underscores this observation. 
Second, the temperature-dependent magnetization data in Fig.~\ref{flo:Magnetization}(a-c) and (d, inset) clearly show the ferromagnetic (FM) nature of the low-temperature phase: at base temperature there is a spontaneous moment that vanishes above $T_{\rm{C}}$. Ferromagnetism is also indicated by the characteristic field sweeps in panels (j-k).  Note, however, the difference between the field-cooled and zero-field cooled magnetization in Fig. \ref{flo:Magnetization}(c) at 0.02T below the transition temperature, indicating some difference in field-cooled vs. zero-field cooled magnetic order for low fields. For higher fields, (panels a and b), there is no visible difference between fc-fh and zfc-fh data.
Third, the considerable hysteresis in temperature sweeps in Fig.~\ref{flo:Magnetization}(a-c) confirms previous reports of a first-order phase transition in $\rm{Yb_2Ti_2O_7}$ \cite{FirstOrderTransition, Chang}, which occurs discontinuously via nucleation and domain growth, causing significant hysteresis in the order parameter vs. temperature. The first order nature is also confirmed by the spontaneous moment (Fig.~\ref{flo:Magnetization}(d, inset), computed from field-dependent magnetization (as described in supplementary material) having no temperature dependence below $T_{\rm{C}}$.
Fourthly and finally, the field sweeps in Fig.~\ref{flo:Magnetization}(f-k) show asymmetric minor hysteresis loops for temperatures between 0.3~K and 0.4~K (where the field scan passes through the phase boundary twice). 
This hysteresis is an additional indication of the discontinuous first-order phase boundary.

The neutron diffraction measurements in Fig. \ref{flo:NeutronData} clearly show the onset of the magnetic order seen in the magnetization, and corroborate the reentrant phase diagram of $\rm{Yb_2Ti_2O_7}$: the temperature scans in Fig. \ref{flo:NeutronData}(b-d) show transition temperatures (defined as the temperature where the Bragg intensity begins to increase) following the same field-dependence as heat capacity and magnetization. 
Additionally, the data in Fig. \ref{flo:NeutronData}(b-d) confirm the first-order nature of the phase transition, with massive hysteresis in the temperature scans, even though the scans were extremely slow (the scans in panels (c) and (d) had sweep rates of 0.6mK/min). Note, however, that no hysteresis is apparent in the 100mK field sweep of the $(2\bar20)$ peak (Fig. \ref{flo:Magnetization}(a)). This suggests either a second order phase transition, or a weakly first order transition. 

Closer examination of the $(2\bar20)$ neutron diffraction provides more clues about the magnetic order. In particular, the field-dependent scattering in Fig.~\ref{flo:NeutronData}(a) is inconsistent with that of a kagome ice phase. We compared the $\rm{Yb_2Ti_2O_7}$ data to the $(2\bar20)$ scattering for $\rm{Ho_2Ti_2O_7}$ entering the kagome ice phase \cite{Ho2TI2O7_kagice}, which has step-like increases in $(2\bar20)$ intensity signaling entry and exit from the kagome-ice state. For $\rm{Yb_2Ti_2O_7}$, the steady increase in scattering suggests that spins continuously cant from a ferromagnetic ordered state as field increases, until at 0.57 T they undergo a transition to a state polarized along $\langle 111 \rangle$, causing a drop in $(2\bar20)$ intensity.

To determine the low T ordered spin state we collected difference data at 
$(2\bar20)$,  $(4\bar40)$, and $(311)$. We performed a refinement to the observed Bragg intensities using the structures reported by Gaudet et. al. (two canted in, two canted out) \cite{GaudetRoss_order} and  Yaouanc et. al. (all canted in all canted out) \cite{Yaouanc_order}, allowing the canting angle and moment size to vary. More details are provided in the supplementary information. The results are shown in Table \ref{flo:refinement}. Although our refinement contained only three peaks and did not account for extinction, some basic conclusions can be drawn. 
First, we found that fitting peak intensities to either structure requires the existence of ferromagnetic domains. Evidence for ferromagnetic domains was previously observed \cite{Chang}, and the presence of domains is consistent with the vanishing zero field magnetization in Fig. \ref{flo:Magnetization}(f-k).
Second, our refined moment and angle are consistent with the Gaudet et. al. structure, but not with the Yaouanc et. al. structure. Given the limited data in our refinement, this should not be taken as conclusive, but as corroborating evidence for the two-in-two-out structure.

\begin{table}[h]
\centering
\begin{ruledtabular}
\begin{tabular}{c c c | c c c c}
 Structure & $\mu$ ($\mu_B$) & $\theta$ & $\chi^2$ & $\chi^2_{domain}$ & $\mu_{fit}$ ($\mu_B$) & $\theta_{fit}$ \\
 \hline
 Gaudet\cite{GaudetRoss_order} & $0.90(9)$ & $14(5)^{\circ}$ & 85.5 & 11.66 & $0.90(3)$ & $8(6)^{\circ}$ \\
 Yaouanc\cite{Yaouanc_order}  & $0.95(2)$ & $26.3(6)^{\circ}$  & 85.8 & 18.13 & $0.851(2)$ & $6.2(1)^{\circ}$ \\ 
\end{tabular}
\end{ruledtabular}
\caption{Refinement to neutrons scattering intensities, allowing canting angle and ordered moment size to vary.}
\label{flo:refinement}
\end{table}


\begin{figure}
\centering\includegraphics[scale=0.52]{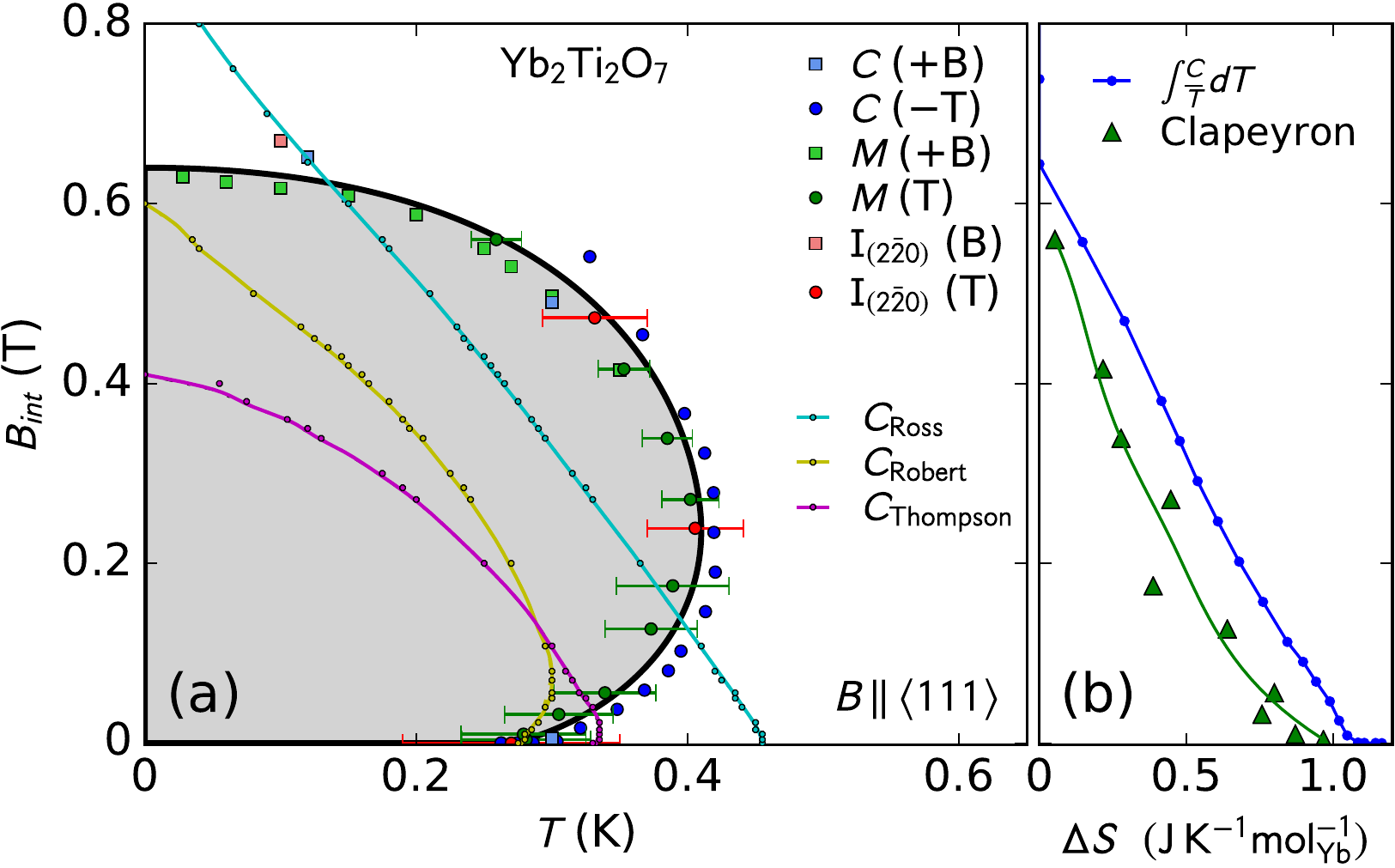}

\caption{(a) Phase diagram of $\rm{Yb_2Ti_2O_7}$ in a $\langle 111 \rangle$ oriented field, built from heat capacity, magnetization, and neutron scattering. Heat capacity points denote peak location (see Fig.~\ref{flo:HeatCapacity}), magnetization points denote inflection points (see Fig.~\ref{flo:Magnetization}), and neutron scattering points denote where intensity begins increasing (see Fig.~\ref{flo:NeutronData}). Error bars indicate the difference in transition temperature upon heating vs. cooling. Theoretically predicted phase boundaries are shown with the small data points which denote the location of simulated heat capacity peaks. The colored lines are guides to the eye. (b) Change in entropy ($\Delta S$) extracted from heat capacity compared to $\Delta S$ computed from the Clapeyron relation. The green line is a guide to the eye.}

\label{flo:PhaseDiagram}
\end{figure}

We can amalgamate the anomalies in heat capacity, magnetization, and neutron scattering to build a phase diagram of $\rm{Yb_2Ti_2O_7}$ in a $\langle 111 \rangle$ oriented field, shown in Fig. \ref{flo:PhaseDiagram}(a). All measurements concur on the phase boundary's location. We double-checked for consistency between the various data sets by computing $\Delta S$ using the Clapeyron equation for a first order phase boundary $\frac{\Delta S}{\Delta M} = - \mu_0 \frac{\partial H}{\partial T}$, and then compared the result to $\Delta S$ computed from heat capacity, shown in Fig. \ref{flo:PhaseDiagram}(b). (See supplemental materials for more details.) The agreement corroborates the first-order nature of the phase boundary. 

Three model spin Hamiltonians have been determined for $\rm{Yb_2Ti_2O_7}$ by Ross et. al. \cite{Ross_Hamiltonian}, Robert et. al. \cite{Robert2015}, and Thompson et. al. \cite{Coldea2017} through neutron scattering measurements, and we used these as the basis for classical Monte Carlo simulations. The specific heat and average magnetization along $\langle 111 \rangle$ were evaluated by measuring thermal averages employing up to $4\times10^5$ samples per spin. The simulations were carried out on a pyrochlore lattice with $N=4L^3$ spins and periodic boundary conditions. Here, $L$ is the number of unit cells along each direction, which varied from 6 to 30 in our simulations. The results shown are for $L = 10$; other simulations confirmed that finite size effects were small away from phase boundaries. More details of the Monte Carlo calculations and results are provided in the supplementary material.  
The overall field and temperature scale of the computed phase boundaries to FM order are in accord with the data, with the Robert et. al. parameters coming the closest. The simulations also predicts a first order phase boundary throughout. However, the marked lobe-like shape of the phase diagram is not reproduced, except for a small bulge predicted by the Hamiltonian parameters of Robert et. al. that is five times too small in temperature. 

There are two obvious potential sources of the discrepancy. Firstly, long-range magnetic dipolar interactions---not included in our simulation---may cause the spins to align more easily under a field. 
Alternatively, the enhancement of magnetic order in a small field may be interpreted as a suppression of magnetic order in zero field relative to the classical MC result. In other words, quantum fluctuations may suppress the zero-field ordering temperature. Various studies have predicted ground state quantum fluctuations from competition between ferromagnetic and antiferromagnetic phases \cite{Robert2015, Jaubert2015, Yan2017}; the fact that the simulations using the Robert et. al. Hamiltonian---which is near the FM-AFM boundary---comes the closest to the observed phase diagram may lend credence to this theory. Given the evidence for monopoles in the paramagnetic phase \cite{Armitage_monopoles_2016, MonopolesConductivity}, it is also worth noting that the non-collinear spin structure in the $\rm{Yb_2Ti_2O_7}$ ordered phase (FM canted 2-in-2-out) does not preclude collective ground state quantum fluctuations: even though the order is ferromagnetic, the ice-rule required for the QSI effective field theory is approximately preserved in the lattice. In that case, the pocket of phase space that opens up between the MC phase boundary and the observed phase boundary could be a finite temperature manifestation of a U(1) quantum spin liquid. Such quantum fluctuations would lower the transition temperature and might persist in the zero--field ground state. Indeed zero field spin fluctuations in  $\rm{Yb_2Ti_2O_7}$ have been found to be extremely broad in energy \cite{GaudetRoss_order}. This is inconsistent with conventional spin waves of the ordered state and points instead to remnant fractionalized excitations of a spin liquid regime.

In summary, we have used stoichiometric single crystals of $\rm{Yb_2Ti_2O_7}$ to reveal a peculiar reentrant phase diagram in a $\langle 111 \rangle$ oriented field, which current model Hamiltonians cannot explain within a classical short range Monte Carlo simulation. The zero-field ordered state is ferromagnetic with domains, the spins seem to polarize along $\langle 111 \rangle$ above an internal field of 0.65~T, and magnetization hysteresis hints at a correlated domain structure. The peculiar decrease in ordering temperature for  $\langle 111 \rangle$ fields below 0.2~T may be a first tangible indication of the proximity of $\rm{Yb_2Ti_2O_7}$ to a quantum spin liquid phase. 

This work was supported through the Institute for Quantum Matter at Johns Hopkins University, by the U.S. Department of Energy, Division of Basic Energy Sciences, Grant DE-FG02-08ER46544. AS and JK were supported through the Gordon and Betty Moore foundation under the EPIQS program GBMF4532. Use of the NCNR facility was supported in part by the National Science Foundation under Agreement No. DMR-1508249. Financial support through DFG TRR80 (From Electronic Correlations to Functionality) is gratefully acknowledged. S.S. and C.D. acknowledge financial support through the TUM graduate school. Finally, we thank the Homewood High Performance Cluster (HHPC) at Johns Hopkins University for computational resources.

\newpage

\section*{Supplemental Material}

\renewcommand{\thefigure}{S\arabic{figure}}
\renewcommand{\theequation}{S.\arabic{equation}}
\renewcommand{\thepage}{S\arabic{page}}  

\renewcommand{\bibnumfmt}[1]{[S#1]}
\renewcommand{\citenumfont}[1]{S#1}

\setcounter{figure}{0}
\setcounter{page}{1}
\setcounter{equation}{0}

This supplemental material explains the details of the experimental and computational methods used to analyze $\rm{Yb_2Ti_2O_7}$ and simulate its magnetic behavior.

\section{Experimental Methods}

\subsection{Crystal Synthesis}

All measurements were performed on two $\rm{Yb_2Ti_2O_7}$ single crystals grown under identical conditions with the traveling-solvent floating zone method described in detail by Arpino et. al. \cite{SeyedPaper_s}. The heat capacity sample was cut from one crystal, and the neutron diffraction and magnetization spheres came from adjacent parts of the other crystal. Both crystals had no noticeable variations in lattice constant or purity between them or along the each crystal.

\subsection{Heat Capacity}
As noted in the main text, our heat capacity measurements were performed mostly using a long-pulse method wherein we applied a long heat pulse and tracked sample temperature as the sample cooled. The data processing was performed with the LongHCPulse software package \cite{MethodsPaper_s}. In this software, heat capacity is computed using the equation
\begin{equation}
C\frac{dT_s}{dt} =  \kappa(T_s - T_b) \frac{m_s}{M_s},
\end{equation}
where $T_s$ is the sample temperature, $T_b$ is the temperature of the dilution refrigerator, $\kappa$ is the thermal conduction between the sample and refrigerator, $m_s$ is the sample mass, and $M_s$ is the sample molar mass. The heat pulses we applied varied between 20 min. and 1 hr., with temperature rises of up to 300\%.
We show only single-slope analysis on cooling curves because the dual-slope analysis of a first-order hysteretic curves produces unphysical double-peak features. In addition, we found that the heating power the PPMS applied to the sample fluctuated somewhat, yielding substantial experimental error for heat capacity computed from heating pulses.
The short-pulse data was taken using the standard PPMS heat capacity routine.
Our sample for these heat capacity measurements was a near-rectangular 1.04 mg $\rm{Yb_2Ti_2O_7}$ plate with dimensions 0.20 mm$\>\times\>$0.94 mm$\>\times\>$0.92 mm, yielding an average demagnetization factor of 0.68 \cite{PrismDemgnetization_s}.
We corrected for demagnetizing fields by first computing sample-independent magnetization ${\bf M}({\bf H}_{int})$ from sphere magnetization, and solving for  ${\bf B}_{int} = \mu_0 {\bf H}_{int}$ numerically with the equation ${\bf H}_{int}={\bf H}_{ext}-D{\bf M}({\bf H}_{int})$.
For all heat capacity data, the heat capacity of the empty sample holder was measured and subtracted.

While the average demagnetizing factor for the rectangular heat capacity sample was 0.68, we calculate that the actual demagnetizing factor varies between 0.23 (at the smallest edge) to 0.81 (at the center of the sample). This inhomogeneity is likely responsible for the two peaks in heat capacity in low fields. In a weak external field (below 0.1 T), the center of the sample still has domain coexistence and no net field, while the edges are already fully magnetized and experience a nonzero net field. Hence a split transition: the central region has the same $T_{c}$ as in the absence of the external field, while the edges have the $T_{c}$ appropriate for the local value of the net field $H_{int}$.

The estimated field inhomogeneity of approximately 0.05 T can also be translated into a broadening of the transition temperature with the aid of the phase-boundary slope $dH/dT_{c}$, which is of the order of 1 T/K (see Fig. 4). At low fields we thus expect the broadening of the transition on the order of 0.05 K, in agreement with the observations (Fig. 1). 
At $\mu_0 H_{int}=0.23\>$T, $dH/dT_{c}=0$ and we expect peak broadening due to field inhomogeneity of only 0.004 K, less than what is observed. Therefore, at higher fields an additional broadening effect seems to be present.

\subsection{Magnetization}
The magnetization of $\rm{Yb_2Ti_2O_7}$ was measured by means of a bespoke vibrating coil magnetometer (VCM) as combined with a TL400 Oxford Instruments top-loading dilution refrigerator \cite{Legl2010a_s,Krey2012_s,Legl2012_s, NIST_disclaimer_s}. As its main advantage our VCM offers excellent thermal coupling without risk of mechanical vibrations with respect to the applied magnetic field, which is highly homogeneous. This contrasts Faraday magnetometers, in which the sample is exposed to a field gradient, or extraction magnetometers, where the sample is moved with respect to the field, either of which may generate parasitic signal contributions due to uncontrolled field and temperature histories. 

Data were recorded at temperatures down to $\sim0.028\>$K under magnetic fields up to 5 T at a low excitation frequency of $19\>$Hz and a small excitation amplitude of $\sim0.5\>$mm. The sample temperature was measured with a $\rm RuO_2$ sensor mounted next to the sample and additionally monitored with a calibrated Lakeshore $\rm RuO_2$ sensor attached to the mixing chamber in the zero-field region.

As noted in the text, the sample for magnetization was a 4.7 mm sphere. To suspend the sample in the VCM it was glued with GE varnish into an oxygen-free Cu sample holder composed of two matching sections fitting accurately the size of the sphere (see Fig. \ref{flo:SampleMount_s}). The sample holder with the sample mounted was firmly bolted into a Cu tail attached to the mixing chamber of the dilution refrigerator. This provided excellent thermal anchoring of the sample across the entire surface of the sphere, while keeping its position rigidly fixed without exerting strain \cite{Legl2010a_s}. The sample holder is shifted with respect to the axis of the cold finger in a way that the sample is centred on the cold finger axis and therefore also on the axis of the VCM coil.




\begin{figure}
\centering\includegraphics[scale=1.0]{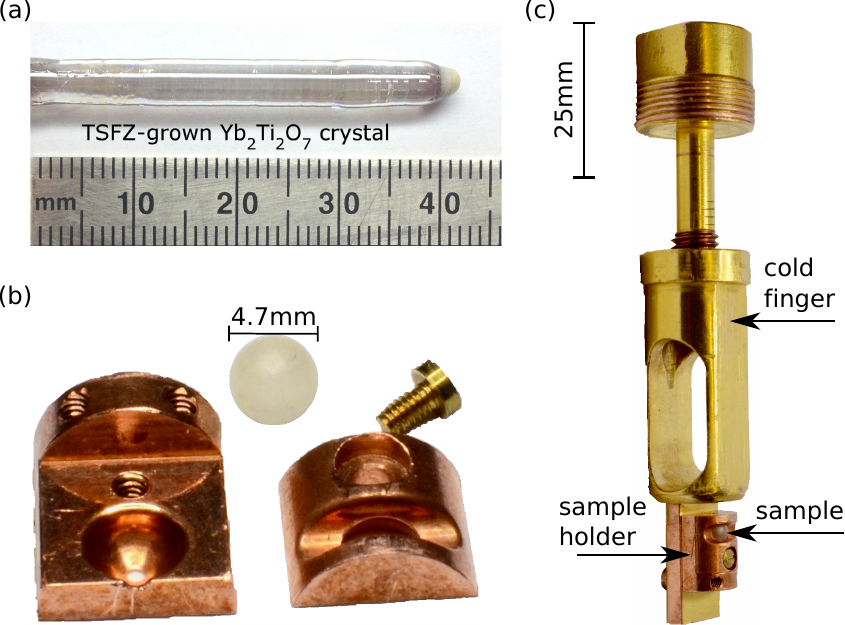}

\caption{(a) TSFZ technique (solvent = 30wt\% rutile Ti02 and 70wt\% $\rm{Yb_2Ti_2O_7}$) produces a large single crystal of $\rm{Yb_2Ti_2O_7}$ that is clear and colourless (image taken from Arpino et al. \cite{SeyedPaper_s}). (b) Spherical sample ground from the stoichiometric single crystal and the oxygen-free Cu sample holder composed of two matching sections fitting accurately the size of the sphere. (c) Sample holder mounted on the cold finger which is then bolted into the Cu tail attached to the mixing chamber of the dilution refrigerator.}

\label{flo:SampleMount}
\end{figure}

To determine the signal contribution of the sample, the empty sample holder was remeasured and subtracted. The signal of the empty sample holder was found to be small with a highly reproducible field dependence and an essentially negligible temperature dependence. The signal of the sample was calibrated quantitatively at $2\>$K and $3\>$K against the magnetization measured in a Quantum Design physical properties measurement system determined also at $2\>$K and $3\>$K, as well as a Ni standard measured separately in the VCM \cite{Legl2010_s}.

To measure temperature dependence, three procedures were used: (i) After cooling at zero magnetic field from $\sim\,$1 K, the magnetic field was applied at base temperature and data collected while heating continuously at a rate of 5 mK/min. This is referred to as zero-field-cooled / field-heated (zfc-fh). (ii) Data were recorded while cooling in the same unchanged applied magnetic field. This is referred to as field-cooled (fc) (iii) After initially cooling in the applied magnetic field, data were recorded while heating continuously at a rate of 5~mK/min in the same unchanged magnetic field. These data are referred to as field-cooled / field-heated (fc-fh). These data were collected in applied fields of $\mu_0 H_{ext} =$ 900, 600, 450, 370, 300, 200, 150, 75, 50, 20, and 10 mT; the 20, 300, and 600 mT data are shown in Fig. 2.
Similarly, the magnetic field dependence data were collected using three different procedures: (iv) After zero-field-cooling a field sweep from $0\,\rightarrow\,$+1~T, denoted (A1). (v) A field sweep +1T$\,\rightarrow\,$-1T, denoted (A2). (vi) A related field sweep from -1T$\,\rightarrow\,$+1T, denoted (A3). For temperatures above 0.05~K all data were recorded while sweeping the field continuously at 15~mT/min, whereas measurements at 0.028~K, the lowest temperature accessible, were carried out at a continuous sweep rate of 1.5~mT/min to minimize eddy current heating of the Cu mount.

The spontaneous magnetization $M_0$ shown in the inset in Fig. 2(d) was obtained by extrapolating the low field behaviour of the zero-field-cooled magnetization data linearly (Fig. 2(d)) to zero field. The black and red lines are the transition temperatures $T_{\rm{C}}$ determined by specific heat and magnetization measurements, respectively. The red shaded area indicates the difference in $T_{\rm{C}}$ upon heating vs cooling in the temperature dependence of the magnetization. $M_0$ remains almost constant at $\sim\,0.62\,\mu_{\rm{B}}\rm{Yb}^{-1}$ before vanishing at $T_{\rm{C}}$. The lack of a temperature dependence of $M_0$ below $T_{\rm{C}}$ further supports the first-order nature of the phase transition.

The remnant magnetization of the superconducting magnet in zero-field can be estimated from the hysteresis loops of Fig. 2 \cite{Krey2012_s}. When we look at the hysteresis loop taken at 900mK, where the hysteresis is closed, we find at $M=0$ a difference in up- and down-sweep of <10mT. This is an instrumental offset of the VCM. Because of this, we can be confident that the zero-field cooled data had a remnant field of <10mT, well within the range where the internal magnetic field is zero.

\subsubsection*{Relating Heat Capacity to Magnetization}

\begin{figure}
\centering\includegraphics[scale=0.5]{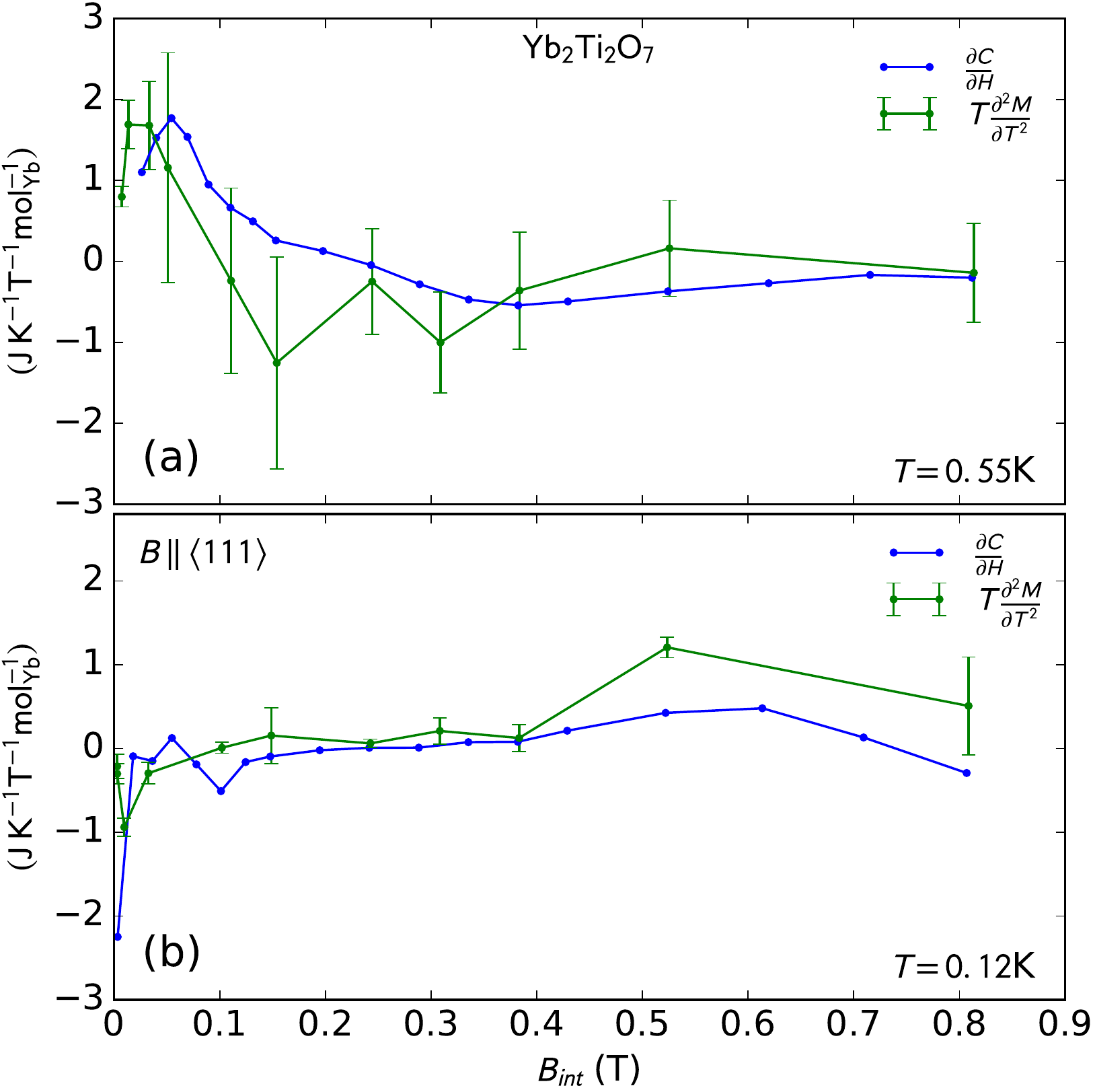}

\caption{Calculation relating $C$ to $M$ with $T\frac{\partial^{2}M}{\partial T^{2}}=\frac{\partial C}{\partial H}$ at (a) 0.55 K and (b) 0.12 K. Error bars indicate one standard deviation.}

\label{flo:CMrelation}
\end{figure}

As a consistency check, we can relate the heat capacity data to the magnetization data using the expression $T\frac{\partial^{2}M}{\partial T^{2}}=\frac{\partial C}{\partial H}$ for temperature ranges where our system is in equilibrium. The results for $T=0.55\>{\rm K}$ and $T=0.12\>{\rm K}$ (away from the first-order transition) are shown in Fig. \ref{flo:CMrelation}.  The large error bars are from taking a numerical second derivative of slightly noisy data (uncertainty was estimated by computing the numerical derivative with combinations of data points between $n-3$ and $n+3$ and taking the standard error of the mean). Even though these measurements were taken on two different samples with different geometries in different instruments, the relation between them holds to within uncertainty (with the exception of 0.52 T at 0.12 K, which is near a phase boundary and not in equilibrium), and $C$ relates to $M$ as expected.

\subsection{Neutron Diffraction}
We collected neutron diffraction data using the SPINS triple axis spectrometer at the NCNR with $E_i=E_f=5 \> {\rm meV}$ neutrons. We used a beryllium filter before the analyzer, 40' collimators before and after the sample, and a flat analyzer. Using the ResLib software package \cite{ResLib_s}, we calculate that this configuration gives a full width at half maximum incoherent elastic energy resolution of $0.23 \> {\rm meV}$.

\section{Computational Methods}

\begin{figure*}
\centering\includegraphics[scale=0.6]{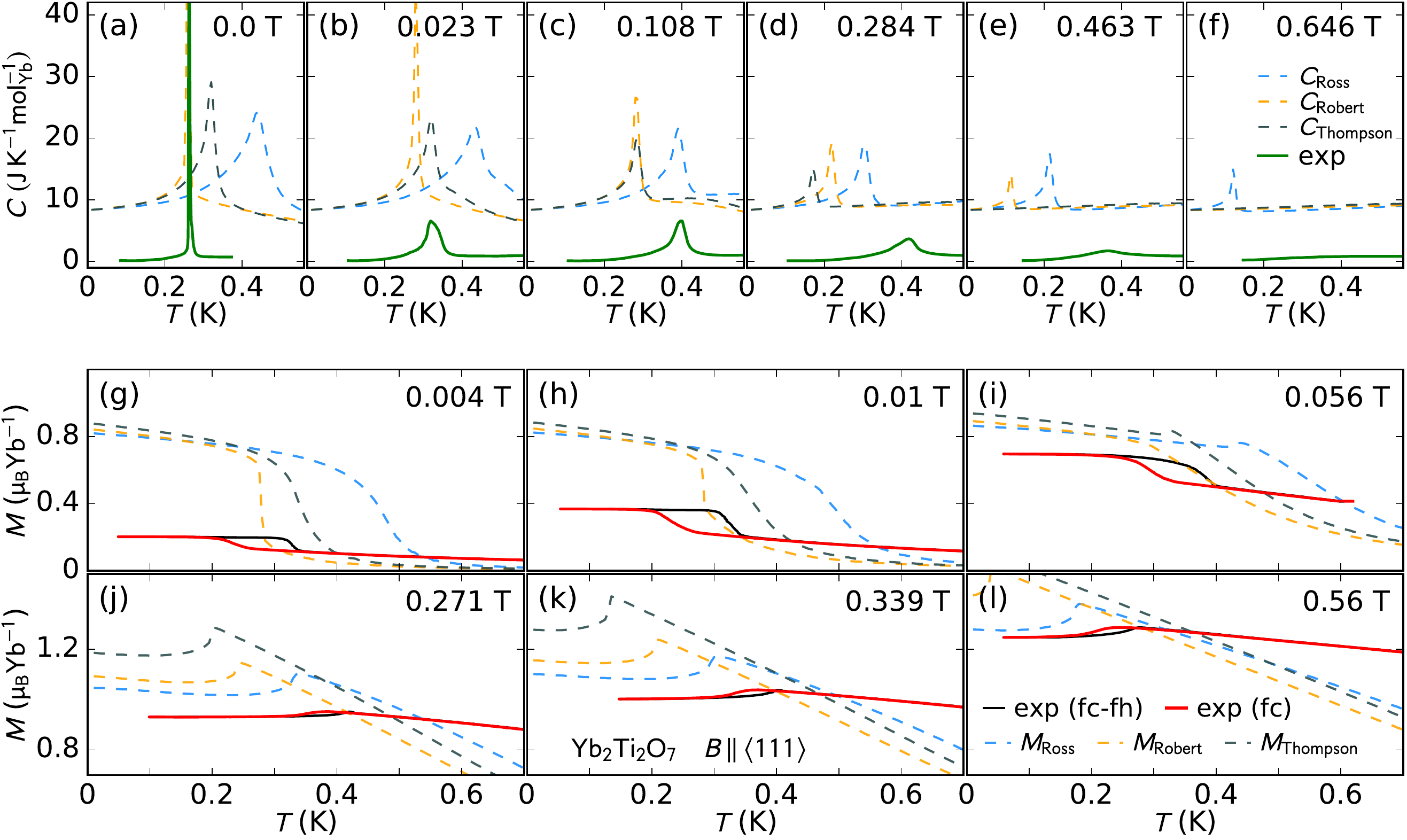}

\caption{Classical Monte-Carlo simulations of the specific heat (for 864 spins) and the magnetization measured along the [1 1 1] direction (for 4000 spins) per spin compared to the corresponding experimental data. (a-f) Heat capacity vs. temperature. (g-l) Magnetization vs. temperature. Fields displayed are internal magnetic fields.}

\label{flo:TheoryVsExperiment}
\end{figure*}

\subsection{Magnetic Structure Refinement}
The magnetic refinement to the neutron diffraction peaks was carried out on three peaks available in our instrumental configuration. In zero-field temperature scans, the $(2\bar20)$, $(4\bar40)$, and $(311)$ peaks change in intensity by $(2.04 \pm 0.06)\%$, $(0.94 \pm 0.06)\%$, and $(4.80 \pm 0.02)\%$, respectively. (Uncertainties indicate one standard deviation.)
We computed the magnetic Bragg intensity by multiplying the computed nuclear structure factor by the observed change in intensity. This was done in order to eliminate the effects of different wavelengths used on the different peaks ($(2\bar20)$ and $(311)$ data were collected with 5 meV neutrons, $(4\bar40)$ data was collected with 20 meV neutrons).

We first refined the structure assuming a net FM moment along the $(100)$ direction with no domains, but the fit was not very good. $\chi^2$ was reduced by 85\% by assuming randomly distributed ferromagnetic domains  along $(100)$, $(\bar100)$, $(010)$, $(0\bar10)$, $(001)$, or $(00\bar1)$, as shown in Table 1.

\subsection{Classical Monte Carlo Simulation}
The pyrochlore lattice comprises sites on tetrahedra whose centers form an FCC lattice, alternately one can view it as made of 4 FCC sublattices whose origins are at (in units of lattice constant $a$) $\vec{r}_0=(1/8,1/8,1/8)$, $\vec{r}_1=(1/8,-1/8,-1/8)$, $\vec{r}_2=(-1/8,1/8,-1/8)$ and $\vec{r}_3=(-1/8,-1/8,1/8)$.

We work with $N=4L^3$ lattice sites; coordinates of vertices on each sublattice are, 
\begin{equation}
	\vec{R}_i(n_1,n_2,n_3) \equiv n_1 \vec{a}_1 + n_2 \vec{a}_2 + n_3 \vec{a}_3 + \vec{r}_i
\end{equation}
where $n_1,n_2,n_3$ are integers from 0 to $L-1$, $\vec{a}_1,\vec{a}_2,\vec{a}_3$ are the primitive 
lattice vectors of the FCC lattice and $i$ is a sublattice index $0,1,2,3$. Periodic boundary conditions were employed along the $\vec{a}_1,\vec{a}_2,\vec{a}_3$ directions. 

In Yb$_2$Ti$_2$O$_7$, the Yb$^{3+}$ ions are best described by the total angular momentum $\mathbf{J} = \mathbf{L} + \mathbf{S}$, because of strong spin-orbit coupling. The crystal field induced by the atoms around the magnetic atom splits the $2J+1$ degeneracy and leads to an effective spin-1/2 ground state (the doublet of ${\bf J}=7/2$) \cite{GaudetRoss_CEF_s}.

The generic form of the effective Hamiltonian in terms of spin degrees of freedom (in global coordinates) that includes the nearest neighbor exchange coupling and magnetic field is,
\begin{equation}
H = \frac{1}{2} \sum_{ij} J^{\mu\nu}_{ij} S^{\mu}_{i} S^{\nu}_{j} - \mu_{B} H^{\mu} \sum_{i} g^{\mu \nu}_{i} S^{\nu}_{i}
\label{eq:Ham}
\end{equation}
The $3\times 3$ interaction matrix $J^{\mu \nu}$ is described completely by four independent parameters $J_1,J_2,J_3,J_4$ and is bond dependent. For example, for the bond connecting sites 0 and 1 the interaction matrix $J_{01}$ is, 
\begin{equation}
\left(\begin{array}{ccc}
 J_2 & J_4 & J_4 \\
-J_4 & J_1 & J_3 \\
-J_4 & J_3 & J_1 \end{array} \right) 
\end{equation}
By rotating to the orientation of the other three bonds, we get the remaining J matrices. Coupling to an external magnetic field is treated as a perturbation with the anisotropic response accounted for by a sublattice dependent $g$ tensor.

We have considered the Hamiltonian in Eq.~\ref{eq:Ham} with parameters provided by Ross et al. \cite{Ross_Hamiltonian_s},
$$
\begin{array}{l l}
	J_1 = -0.09 \;\;\text{meV}\quad & J_2= -0.22 \;\;\text{meV} \\
	J_3= -0.29 \;\;\text{meV} & J_4= +0.01 \;\;\text{meV} \\
        g_{z}= 1.80   &	g_{xy}/g_{z}= 2.4
\end{array} 
$$

parameters provided by Robert et. al. \cite{Robert2015_s},
$$
\begin{array}{l l}
	J_1 = -0.03 \;\;\text{meV}\quad & J_2= -0.32 \;\;\text{meV} \\
	J_3= -0.28 \;\;\text{meV} & J_4= +0.02 \;\;\text{meV} \\
        g_{z}= 2.06   &	g_{xy}= 4.09
\end{array} 
$$
and parameters provided by Thompson et. al. \cite{Coldea2017_s}.
$$
\begin{array}{l l}
	J_1 = -0.028 \;\;\text{meV}\quad & J_2= -0.326 \;\;\text{meV} \\
	J_3= -0.272 \;\;\text{meV} & J_4= +0.049 \;\;\text{meV} \\
        g_{z}= 2.14   &	g_{xy}= 4.17
\end{array} 
$$

Ignoring quantum effects completely, we treat the spins on every site ($S^{x}_i, S^{y}_i, S^{z}_i$) 
as classical vectors of length 1/2. Given a set of spin directions for $N$ lattice sites,
$S \equiv ((S^{x}_1, S^{y}_1, S^{z}_1),(S^{x}_2, S^{y}_2, S^{z}_2)....(S^{x}_N, S^{y}_N, S^{z}_N))$ 
we use Eq.~\ref{eq:Ham} to compute the energy $E(S)$. A standard Metropolis Monte Carlo algorithm is used 
to sample according to the Boltzmann distribution $\exp(-\beta E(S))$ (where $\beta$ is the inverse temperature $1/k_B T$ and $k_B$ is the Boltzmann constant) to compute thermal averages of quantities such as the energy and magnetization. 
The moves in this algorithm correspond to choosing a spin on a site randomly and moving it in a cone around its present direction. The maximum radius of the circle traced out on the sphere is a free parameter that was chosen to keep the acceptance rates for most generic situations to about 50 percent. Other move choices were also employed to either test the accuracy of the simulation or used in conjunction with the conical move to increase its sampling efficiency. 10-50 percent of the initial run time was counted towards equilibration during which no statistics were collected. A sweep consisted of $N$ single spin moves, at the end of which the physical quantity was recorded as part of the averaging procedure; $4 \times 10^5$ - $2 \times 10^6$ such sweeps were used. Different starting configurations did not appear to influence the final measured averages, at least for the smaller systems, indicating adequate equilibration. 

In the text and supplementary information, the results presented are for $L=6$ or $L=10$ ($N=864$ or $N=4000$ sites), and checks were performed to ensure that the finite size effects away from the transition were indeed small, for all three parameter sets. At the critical temperature, the specific heat peak location (peak height) is weakly (strongly) size dependent, as expected. The peak height also depends on the precise values of temperature at which the simulations were performed;  however, the small error (10 mK or less) in the reporting of the peak location does not affect the general conclusions arrived at by our theoretical analysis. 

We also investigated the nature of the transition by carrying out qualitative experiments with up to $L=30$ ($N=108000$ sites) near the phase boundary to establish the tendency of the system to tunnel between two phases. We found that the transition was discontinuous at all fields, albeit weakly, suggesting the first-order nature of the phase transition. It must be noted that a coarse grained model with an effective ferromagnetic exchange and (non-zero) negative cubic anisotropy has a discontinuous phase transition~\cite{Cardybook_s}.

Comparison to experimental data is shown in Fig. \ref{flo:TheoryVsExperiment}. The simulations match the overall temperature and field scales of the phase diagram. 
Note that in low magnetic fields--Fig. \ref{flo:TheoryVsExperiment}(g-h)--the calculated magnetization is much higher than the measured magnetization; this is due to the ferromagnetic domains in the real material, which our simulations did not include.


\subsection{$\Delta S$ from the Clapeyron Relation}

The Clapeyron Equation used in the text is derived as follows.
Across a first order phase boundary, the free energy is equal: $F_1 = F_2$. We write free energy as $dF = -SdT - \mu_0 MdH$, where $M$ is magnetization and $H$ is the magnetic field. On a phase boundary, $H_1 = H_2$ and $T_1 = T_2$. From this, one derives
\begin{equation}
\frac{\Delta S}{\Delta M} = - \mu_0 \frac{\partial H}{\partial T}.
\end{equation}
$\Delta S$ across the phase boundary was computed using the Clapeyron relation by taking $\frac{\partial H}{\partial T}$ from a smooth curve chosen by eye to best fit the phase boundary data (shown as a grey shape in Fig. 4[a]), and $\Delta M$ to be the difference in temperature-dependent magnetization at $T_c$ between the heating and cooling curves in the magnetization ($T_c$ is the average temperature of the inflection points in magnetization upon heating and cooling--see Fig. 2).

As a comparison, $\Delta S$ was computed from heat capacity data by integrating $C/T$ over the heat capacity peak. To isolate the peak, we subtracted the heat capacity taken at $B_{ext}=0.8$ T (the lowest field outside the phase boundary) and integrated from 0.2 K to 0.5 K. This procedure approximately isolated the heat capacity from the transition, enabling a comparison to the calculation of $\Delta S$ from the Clapeyron relation. As is shown in Fig. 4(b), the two methods of computing  $\Delta S$ agree quite well.


\begin{thebibliography}{33}
\bibitem{NIST_disclaimer}Certain commercial instruments are identified in this pa-
per to foster understanding. Such identification does not
imply recommendation or endorsement by the National
Institute of Standards and Technology, nor does it imply
that the instruments identified are necessarily the best
available for the purpose.
\end{thebibliography}
\begin{thebibliography}{14}
\bibitem{NIST_disclaimer_s}Certain commercial instruments are identified in this paper to foster understanding. Such identification does not imply recommendation or endorsement by the National Institute of Standards and Technology, nor does it imply that the instruments identified are necessarily the best available for the purpose.
\end{thebibliography}
\end{document}